# Local-Sensitive Connectivity Filter (LS-CF): A Post-Processing Unsupervised Improvement of the Frangi, Hessian and Vesselness Filters for Multimodal Vessel Segmentation

**Erick O. Rodrigues [1,*], Lucas O. Rodrigues [2], João H. P. Machado [1], Dalcimar Casanova [1], Marcelo Teixeira [1], Jeferson T. Oliva [1], Giovani Bernardes [3] and Panos Liatsis [4]**

1 Department of Academic Informatics (DAINF), Universidade Tecnologica Federal do Parana (UTFPR), Pato Branco 85503-390, PR, Brazil
2 Graduate Program of Sciences Applied to Health Products, Universidade Federal Fluminense (UFF), Niteroi 24241-000, RJ, Brazil
3 Institute of Technological Sciences (ICT), Universidade Federal de Itajuba (UNIFEI), Itabira 35903-087, MG, Brazil
4 Department of Electrical Engineering and Computer Science, Khalifa University of Science and Technology, Abu Dhabi P.O. Box 127788, United Arab Emirates
* Correspondence: erickrodrigues@utfpr.edu.br

**Abstract:** A retinal vessel analysis is a procedure that can be used as an assessment of risks to the eye. This work proposes an unsupervised multimodal approach that improves the response of the Frangi filter, enabling automatic vessel segmentation. We propose a filter that computes pixel-level vessel continuity while introducing a local tolerance heuristic to fill in vessel discontinuities produced by the Frangi response. This proposal, called the local-sensitive connectivity filter (LS-CF), is compared against a naive connectivity filter to the baseline thresholded Frangi filter response and to the naive connectivity filter response in combination with the morphological closing and to the current approaches in the literature. The proposal was able to achieve competitive results in a variety of multimodal datasets. It was robust enough to outperform all the state-of-the-art approaches in the literature for the OSIRIX angiographic dataset in terms of accuracy and 4 out of 5 works in the case of the IOSTAR dataset while also outperforming several works in the case of the DRIVE and STARE datasets and 6 out of 10 in the CHASE-DB dataset. For the CHASE-DB, it also outperformed all the state-of-the-art unsupervised methods.

## 1. Introduction

The retina is a multi-layered tissue of light-sensitive cells that surrounds the posterior cavity of the eye where light rays are converted into electrical neural signals for interpretation by the brain. One of the most important retinal diseases is diabetes, which can result in blindness [1].

A retinal vessel analysis is a medical procedure that can be used as an assessment of risks to the eye and other organs. Retinal vessels contain some of the narrowest vessels of the human body and its analysis can lead to the perception of blood vessel damage caused by early stage metabolic diseases, such as diabetes. It is also possible to estimate the progression of diabetic retinopathy by measuring the rate of dilation of the vessels in response to a flickering light [2].

Fundus imaging is one of the primary methods of screening for retinopathy. Recent advances in digital imaging and image processing resulted in the widespread use of computerized image analysis techniques in all areas of medical sciences, including ophthalmology. Several parameters can be obtained from the retinal vessel structure, such as changes in the thickness of the vessels, the curvature of the vessel structure, and the arteriolar-to-venular ratio (AVR) [1].

These parameters can be computed after an adequate automatic segmentation of the vessel structure [3], which is what we propose in this work. In practice, the vasculature of the eye is still manually segmented or measured by experts, which is a mundane and laborious task. Moreover, segmentation requires expertise and a considerable degree of attention and time. However, even in the case of clinical experts, retinal vessel annotation is prone to human errors and subjectiveness, lacking repeatability and reproducibility. The task becomes even harder in the presence of retinal pathologies, e.g., exudates and hemorrhages [4].

In this work, we improve the popular Frangi filter, a popular vessel segmentation approach, by capitalizing on a simple yet powerful concept called connectivity [4]. The usage of connectivity information mimics the momentum of a brush, instead of evaluating pixels on their own. In a previous work, we associated this technique to a coupled region-growing and machine learning [5,6] approach, creating a unique framework for vessel segmentation called ELEMENT. However, as ELEMENT is based on machine learning, it requires training phases that use segmented vessels annotated by specialists. In the case of a shortage of specialists for a certain image modality, vessels cannot be segmented by ELEMENT.

This work, on the other hand, focuses on improving classical image filters, such as the Frangi filter. We propose two main improvements called the connectivity filter (CF) and the local-sensitive connectivity filter (LS-CF), which are not tied to any machine learning concept, being completely unsupervised. The LS-CF is based on the Frangi filter output and on a heuristic. As it works on top of the Frangi filter, the approach is also multimodal.

The heuristic for this proposal consists of associating a score of connectivity to each pixel of the image. First, we obtain a binary thresholded response of the Frangi filter (other thresholded binary vessel images can be used). After the threshold, the Frangi response generates either an overly connected segmentation with a lot of noise or a sparse segmentation with disconnected and invisible vessel branches. These variations can be seen in Figure 1. Neither of these scenarios (b and f) nor the results in-between (c, d and e) are desirable in terms of vessel segmentation, which is a binary adequate response.

The CF and LS-CF receive the thresholded binary Frangi response as input and produce a gray-level response based on the assumption that parts that yield a great connectivity score are most probably the correctly segmented parts of existing vessels. The connectivity score is computed by recursively iterating over connected pixels. We start by walking over the vessel pixels, and the longer we travel over these connected pixels, the greater the score of that respective pixel. If a pixel is connected to 100 other pixels, it will receive a brighter intensity compared to a pixel that is connected to just 10. We later use this information to improve the disconnections of vessels. Pixels that are not segmented as vessels that are closer to highly connected vessels' pixels (brighter regions) are more likely to be vessels and are more likely to have been misclassified as a non-vessel.

The next section covers the medical aspects of the human eye and its retina, later addressing works in the literature that proposed automated vessel segmentation. Section 3 covers the proposed methodology, where we describe the two newly proposed unsupervised algorithms. Section 4 shows the obtained visual and numerical results, and Section 5 addresses the conclusions and future work.

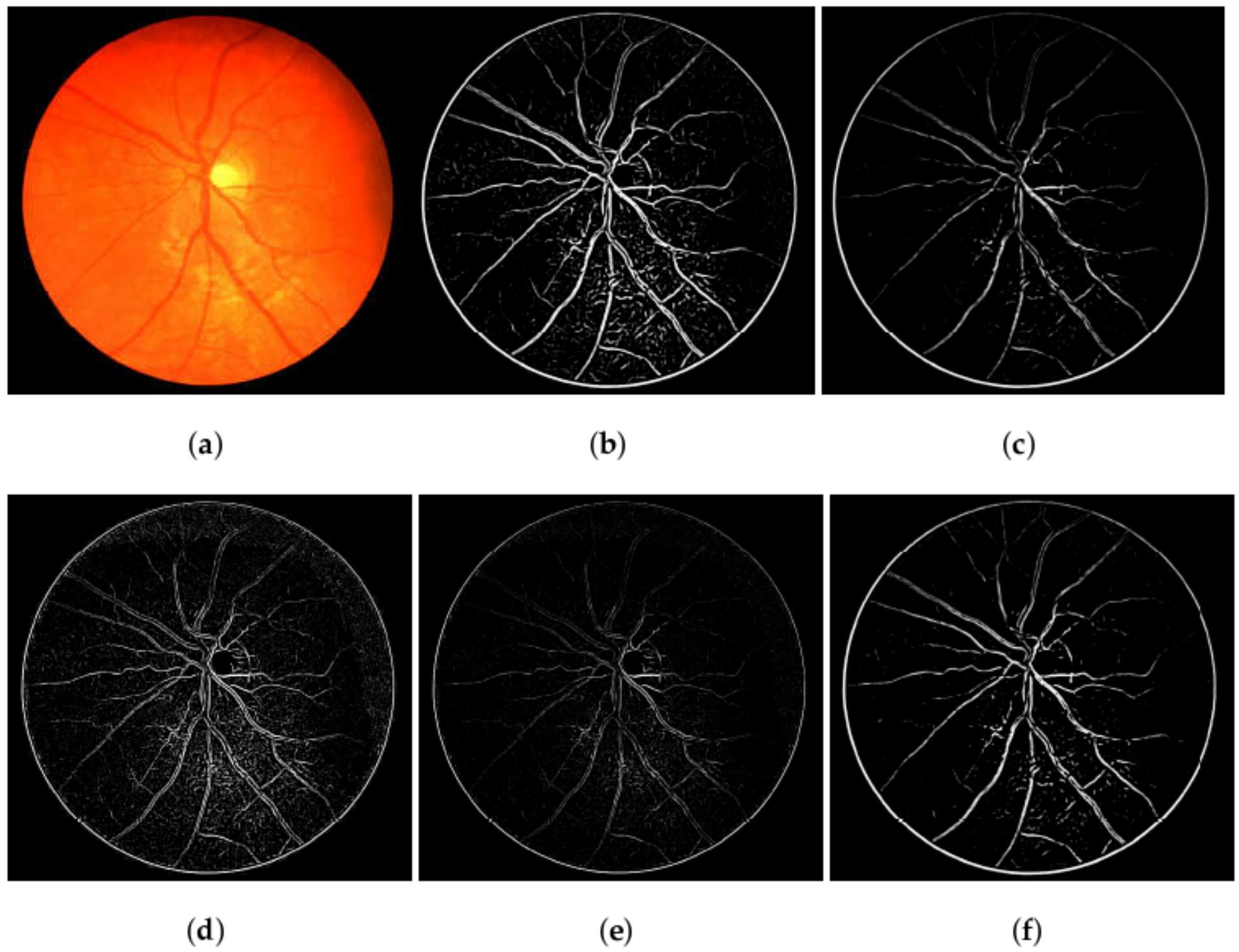

**Figure 1.** Overall variations of the Frangi filter gray-level response. (**a**) Image 11L from the CHASE-DB dataset. (**b**) Thicker vessel caliber response, which suits to the reality of the vessel, but contains several artefacts in non-vessel areas. (**c**) Smoother, less-noisy overall response but erases parts of important vessel branches. (**d**) Very noisy response that highlights most parts of the vessels but is more responsive to the edges of the vessel branches in contrast to their interior. (**e**) Less-noisy version of -e- but still with the same vessel caliber issue. (**f**) Less noisy version of -b- but looses more vessel information.

## 2. Literature Review

A recent study [7] observed that the flicker-induced dilatation of the retinal arterioles is significantly reduced in patients with chronic heart failure compared to patients with cardiovascular risk factors and healthy controls. Computer-based segmentations can also assist with locating and identifying stenoses within the retinal or cardiac vessel structure [8]. Furthermore, the arterial dilation is shown to be weaker in patients with Alzheimer's disease when compared to a control group. Patients with a mild cognitive impairment also show weaker dilation responses but not as significant as with Alzheimer's [9].

The retinal microcirculation is earlier affected in the presence of atherosclerosis, and the retinal vessel caliber is an emerging cardiovascular risk factor. Obesity is also associated with vascular dysfunction. The mean arteriolar-to-venular diameter ratio is impaired in obese subjects when compared to lean cases. Overall, cardiovascular evaluations are associated to the response of the retinal vessels [10].

Stergiopulos et al. [11] focused on modeling the blood flow within the vessels to automatically locate and quantify the degree of severity of stenoses. However, prior to this modeling, vessels were manually segmented. All these highlighted health markers justify concentrating efforts toward automatic vessel segmentation approaches.

The Frangi filter is a popular vessel-enhancing technique proposed in [12,13] that works with a broad extension of modalities, such as X-ray and retinal fundus images. The Frangi filter is based on the Hessian matrix [14]. As defined by the authors [12], their approach searches for tubular geometrical structures. The method uses the eigenvalues of the Hessian matrix to locally determine the likelihood of a vessel. Similar filters are also called Hessian filters in the literature [15,16].

Approaches to the vessel segmentation problem can be divided into two general categories: (1) unsupervised, the ones that do not require training data, and in most cases, these are subsequent applications of image filters; and (2) supervised, which require training information and use machine learning techniques. Usually, well-constructed supervised learning methods that use the response of filters, such as the Frangi, as input will more frequently outperform pure image processing techniques [4].

Unsupervised approaches are frequently based on the Hessian matrix or Frangi [17,18]. However, they can also use mathematical morphology [5,19], region growing, Fourier transform [20], etc. Some methods in the literature are based on deep learning approaches [21]. Some also combine classical filters with supervised learning, usually yielding the best results [4,22–24].

This work focuses primarily on unsupervised techniques, as they do not require training data and can be used as a feature in machine learning approaches. The state-of-the-art and most recent unsupervised approaches in the literature are the works of Memari et al. [25], Tavakoli et al. [26] and Mahapatra et al. [27].

Memari et al. [25] used a complex combination of several techniques: image preprocessing (median filtering, CLAHE), mathematical morphology, the Gabor filter, Frangi filter, level set and a clustering algorithm. Tavakoli et al. [26] used Radon transform and mathematical morphology. Mahapatra et al. [27] also used the Frangi filter, associated to a clustering algorithm. The Frangi filter is prominent in the state of the art when it comes to unsupervised techniques (and also in supervised cases, such as in [4]).

Although the approach of Memari et al. [25] used the level set as one of their steps, which reminds one of the proposal of this work and iterates over the pixel intensities, none of these three state-of-the-art approaches is based on the connectivity of pixels. The first connectivity image filter in the literature is proposed in this work. Although we do use it with vessels, the connectivity filter can also be applied to other application domains, such as a feature descriptor, by adapting the idea to gray-level and color images.

Furthermore, these three state-of-the-art articles primarily focus on the DRIVE and STARE datasets. In this work, we test our proposal with the DRIVE, STARE, CHASE-DB and IOSTAR public datasets, and we also test the performance of the approach on X-ray angiograms (OSIRIX), which is a completely different imaging technique. One major contribution of this work is the multimodal aspect that is confirmed by our experiments. Our proposal also outperformed the state of the art [25,26] for the CHASE-DB dataset. The proposed approach is also cleaner than the approach of Memari et al. [25] that used a combination of several techniques, which leads to several adjustments of parameters and difficulties in replication.

## 3. Proposed Methodology

The methodology pipeline consists of converting colored images to gray images using the green channel, which is commonplace in the literature. Later, we apply the Frangi filter to the image using the MATLAB implementation. The chosen input parameters for the Frangi algorithm vary a little bit across different modalities and were adjusted empirically (per modality) to return acceptable responses (average of 5 combinations of parameters per dataset). We did not overly adjust the parameters to avoid overfitting to a certain modality. The used parameters can be found at [28]. Figure 2a shows an example of the input image and its ground-truth pair, extracted from the DRIVE dataset.

We subsequently apply a threshold operation using the gray-level 100, which was empirical, chosen by visual subjective observation of 10 samples. This number could be chosen based on an optimization of the accuracy. However, we discarded this idea as this could lead to overfitting to the image data that we have, where the methodology could not generalize well to other images. We used this same threshold for all modalities and images in this work. To sum up, if the pixel value is greater than 100, then it is painted white. Otherwise, it is painted black. The result of this threshold operation is shown in Figure 3b.

At last, the proposed connectivity filter is applied to the previous thresholded response. The connectivity filter receives as input a binary image and produces as output a gray-level image containing the connectivity scores for each connected component of the image (Figure 4a). To generate the final segmentation mask, we apply a threshold operation for gray levels greater than 1 (shown in Figure 4b). If we set this threshold to 0, the entire image is painted white. If we set it to 1, it paints white all the pixels that are associated to a connectivity score, irrespective if it is a low or a high connectivity score. A single pixel that is isolated from every other vessel pixel has connectivity score 1, so they are painted white as well. When we select the gray levels greater than 1, we remove these pixels that are noise and are separated from other vessels. This threshold can be adjusted for different segmentation scenarios. Figure 4b represents the final segmentation result for the input image shown in Figure 2.

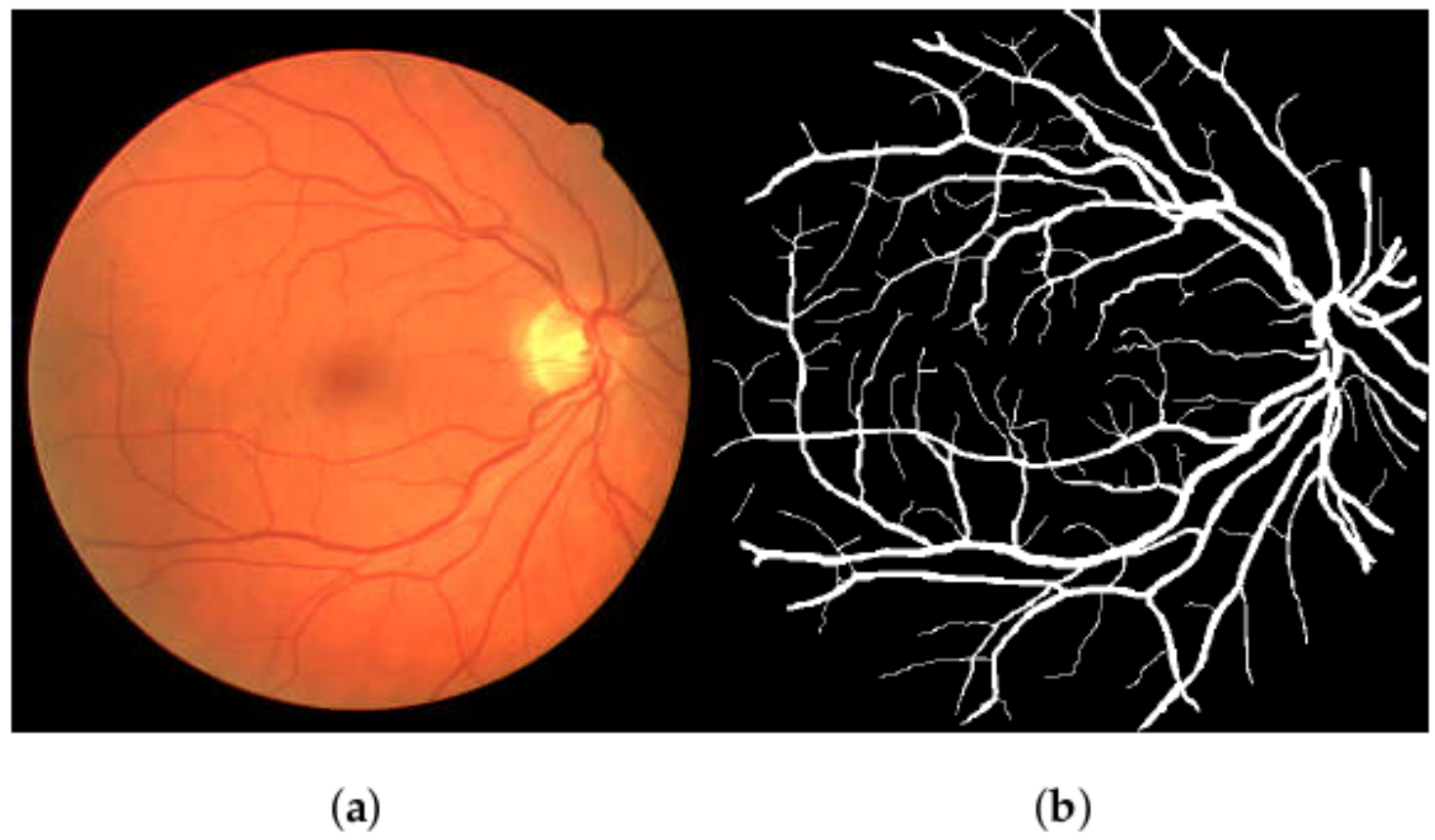

(**a**) (**b**)

**Figure 2.** Image 02 from the test folder of the DRIVE dataset (**a**) and its ground truth (**b**). (**a**) Input image. (**b**) Ground truth.

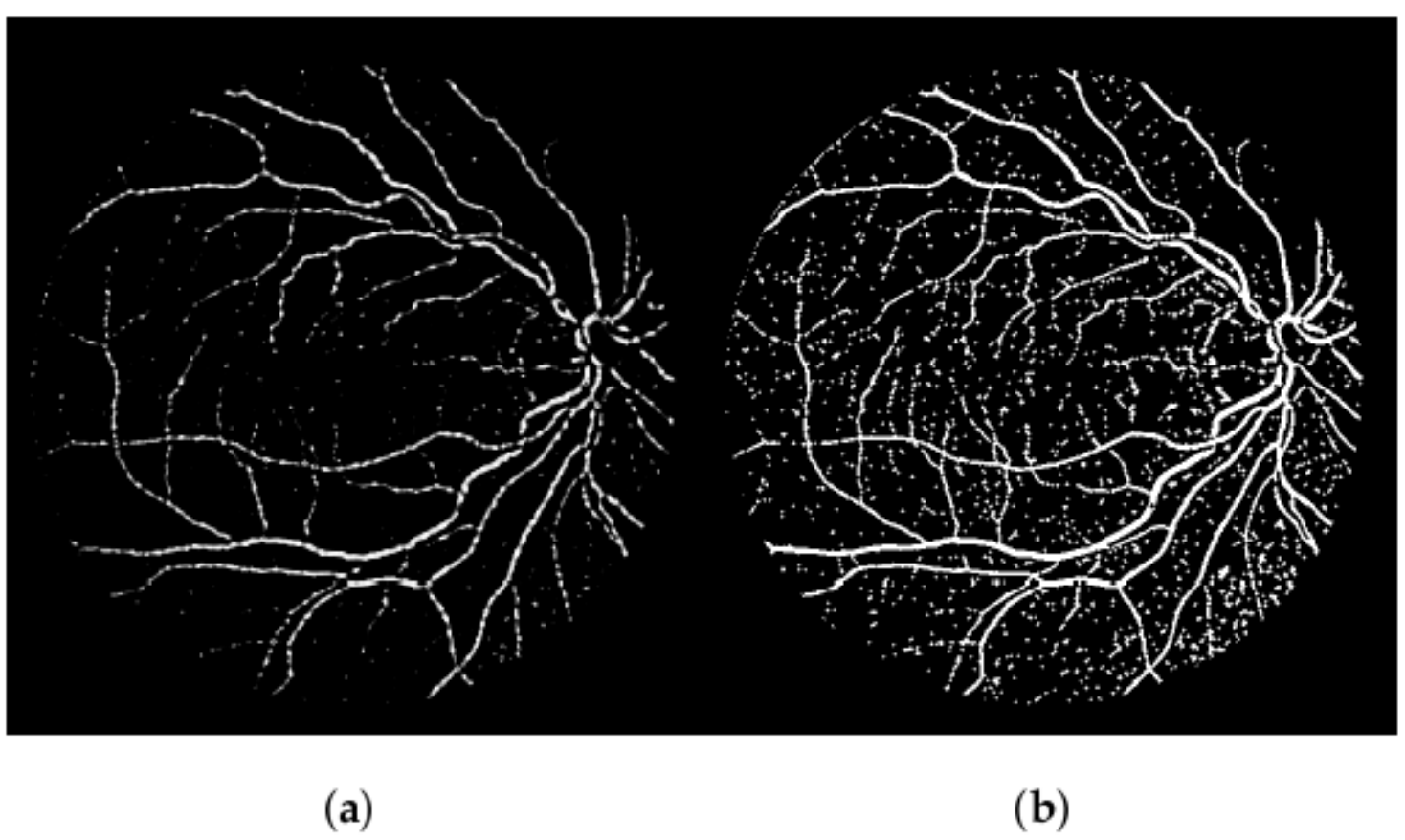

(**a**) (**b**)

**Figure 3.** The Frangi filter response (**a**) and its thresholded version at intensity 100 (**b**). (**a**) Frangi filter response. (**b**) Frangi filter response after threshold.

Figure 4a clearly shows that the highly connected components appear brighter (with a larger associated connectivity score). Some parts of the vessels shown in Figure 4b cannot be seen in Figure 4a because their gray-level value is close to 0 (black), but these elements are there. Figure 4b shows the main vessel branches but also shows a lot of disconnections. To improve this result further, we propose the more intelligent local-sensitive connectivity filter. At total, we propose two algorithms in this work: CF and its improved version called LS-CF.

The connectivity filter (CF) shown in Figure 4a can be represented as a recursive function that iterates over the neighborhood of a pixel, just like a flood-filling approach, while also keeping and incrementing a connectivity associated to that pixel. To obtain the score for a certain pixel, the function *scorePixel* shown in Algorithm 1 is triggered. As the idea is to spread the movement as a flood filling, a binary matrix can be used to mark visited pixels in order to avoid recursive repetitions.

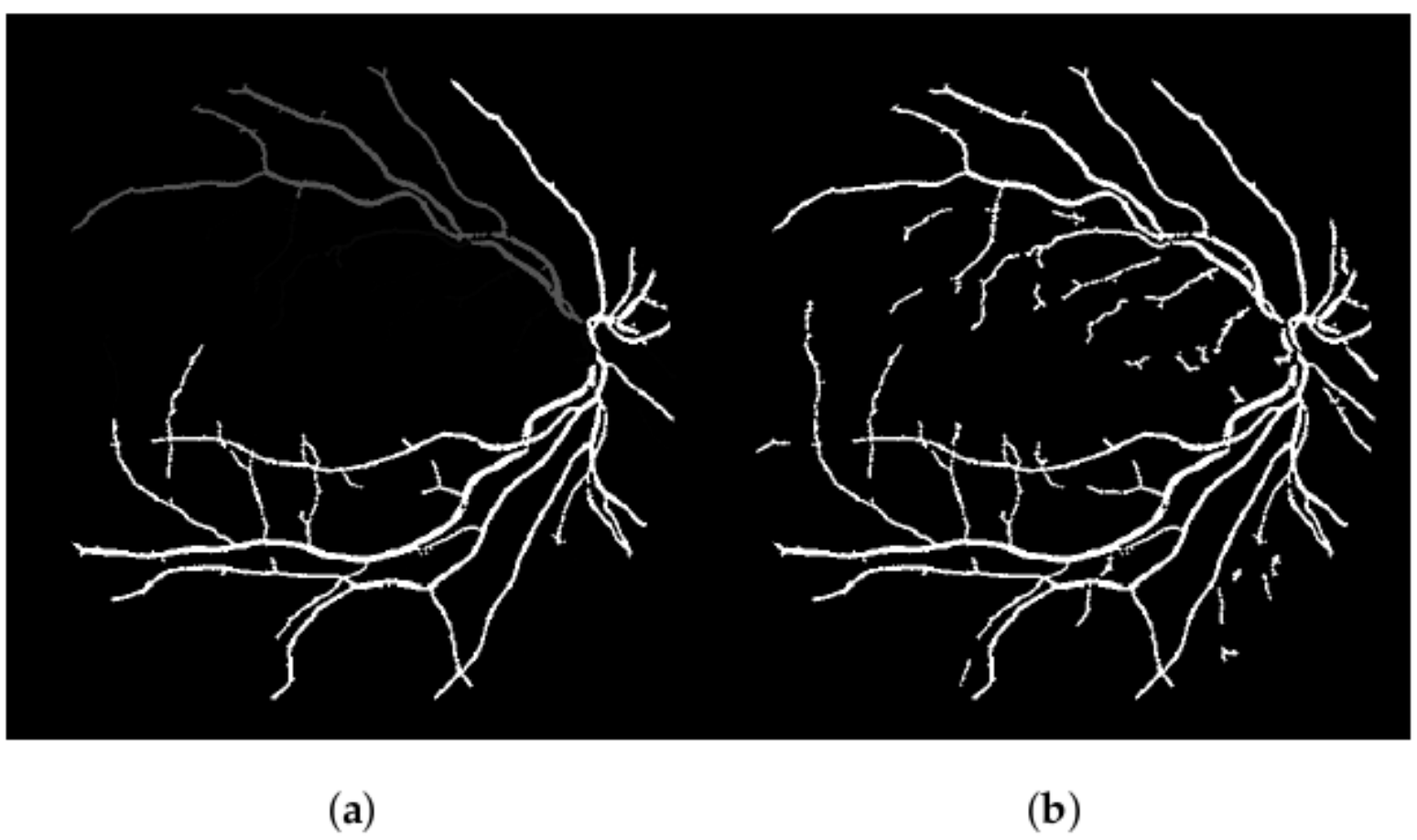

(**a**) (**b**)

**Figure 4.** The connectivity filter output and its thresholded version. (**a**) Connectivity filter. (**b**) Thresholded version ($t > 0$) of (**a**).

---

**Algorithm 1** Connectivity Filter (CF): The recursive flood-filling function that calculates the connectivity scores and outputs the gray-level connectivity image response.

---

**Input:** $I$ representing an image where $I(x, y)$ returns the pixel value at position $(x, y)$, a threshold value $T$, a global (out of the scope of the recursive function) *score* variable for each pixel and a initial chosen pixel position $(x, y)$.

**Output:** The connectivity score stored to the variable *score* for the respective pixel $(x, y)$.

```
initialization;
Function scorePixel(x, y, score):
    if I(x,y) > T and position (x,y) has not been visited then
        score++;
        Mark (x,y) as visited;
    else
        return ;            // terminates the function and the recursion
    end
    for each pixel (j,i) that is at distance 1 from (x,y) do
        scorePixel(j, i, score);
    end
```

---

The CF algorithm, although recursive, was capable of calculating the connectivity scores for all the images used in this work in less than 1 min (38 s in avg for one image of the DRIVE dataset). Algorithm 1 outputs the gray-level image shown in Figure 4a. A threshold must be further applied to the gray-level response to obtain the binary segmentation mask shown in Figure 4b.

The issue with this naive connectivity filter is clear vessel disconnections. Therefore, we modify this naive CF approach, introducing a tolerance score. Let us consider a pixel that has not been marked as vessel; the idea is to iterate around this pixel while incrementing a tolerance score for each visited neighboring pixel. The idea is to tolerate some error and force the neighborhood iteration to continue. If the surrounding pixel is not a vessel, the connectivity score is incremented until a certain pre-defined tolerance.

The result of this modified version, called LS-CF, is shown in Figure 5. The pixels in green show what has changed in comparison to the naive CF approach shown in Algorithm 1. This version is sensitive to local connections and is called local-sensitive connectivity filter (LS-CF).

The great advantage of this heuristic is that it reconnects the vessel branches in any orientation while preserving the vessel thickness, without losing caliber and shape information. Algorithm 2 shows the LS-CF pseudo-code. The algorithm receives a distance to control the neighborhood iteration when computing the tolerance score. Furthermore, it also receives two constants $MAX_{SCORE}$ and $MAX_{DIST}$. The $MAX_{DIST}$ is not strictly necessary, but its usage can reduce the time required for computation. Changes on both of these constants can alter the final result.

---

**Algorithm 2** Local-Sensitive Connectivity Filter (LS-CF): The local-sensitive version of the connectivity filter.

---

**Input:** $I$ representing an image where $I(x,y)$ returns the pixel value at position $(x,y)$, a threshold value $T$, global *connectivityScore* and *toleranceScore* variables for each pixel at position $(a,b)$, a $MAX_{SCORE}$ and $MAX_{DIST}$ constants. All variables $found_d$ start off being set to false.
**Output:** The connectivity score stored to the variable $connectivityScore$ for the respective pixel $(x,y)$.

```
initialization;
Function scorePixel(x, y, connectivityScore, toleranceScore):
    if I(x,y) > T and position (x,y) has not been visited then
        toleranceScore = 0;  // reset once a new connection has been found
        connectivityScore++;
        Mark (x,y) as visited;
        for each position (j,i) at distance 1 from (x,y) do
            scorePixel(j, i, connectivityScore, toleranceScore);
        end
    else if toleranceScore < MAX_SCORE and (x,y) has not been visited then
        toleranceScore++;
        for each neighbor pixel (j,i) at distance d from central pixel (x,y) in ascending
          distance order (first iteration d = 1, then d = 2, ...); as long as
          d < MAX_DIST && found_d == false do
            if I(j,i) > T then
                scorePixel(j, i, connectivityScore, toleranceScore);
                found_d = true;
            else
                toleranceScore++;
            end
            d++;
        end
    else
        return;
    end
```

---

For the neighborhood iterations in Algorithms 1 and 2, we used the Rodrigues distance [29] shown in Equation (1) ($w_1 = w_2 = p = 1$), as it is easy to iterate over the pixel neighborhood iteratively. The algorithm for the discrete iteration can be found in [29]. However, other distances can be used as well, and they will return different segmentation results [29,30].

$$d_{w_1,w_2,p}(x,y) = w_1 \sqrt[p]{\sum_{i=1}^{n} |x_i - y_i|^p} + w_2\ max_{i=1}^{n} |x_i - y_i| \tag{1}$$

The core idea of the algorithm is to travel on top of vessel branch. Eventually, when it encounters non-vessel pixels, the tolerance keeps it going, as it may find vessel pixels in subsequent iterations. If they do, all the pixels that are visited when the algorithm is moving its coordinate are painted as vessel.

It is best to keep the same "momentum" of the stroke. If the algorithm is walking on top of a vessel that is going upward, in each recursive call of the function "scorePixel", we should start visiting pixels that belong to the upward motion. This emulates a brush stroke. The rule is the same for the rest of the orientations.

The main input parameter for this algorithm that can lead to different results are the $MAX_{score}$ and $MAX_{dist}$ variables. These two variables control by how much the filter can "see" future iterations. If we increase both, we tend to have longer run times and more connected pixels. These variables are alike, and they could be written one based on the other, we just separated the concept in two variables to make it more clear, but both control by how much each pixel keeps growing and trying to reconnect to other vessel pixels.

Figure 6 shows the main steps of the LS-CF approach. If we increase the $MAX_{dist}$ parameter as shown in Figure 6e, we obtain more "packed" connections, but we also lose some vessel information, most probably because we obtain higher connectivity scores in this case and, therefore, the noisy disconnected branches get erased after the threshold operation. Figure 6e, when compared to Figure 6c, shows the longer and brighter main vessel branch, horizontally centered, that starts on the top and stops at the bottom of the image.

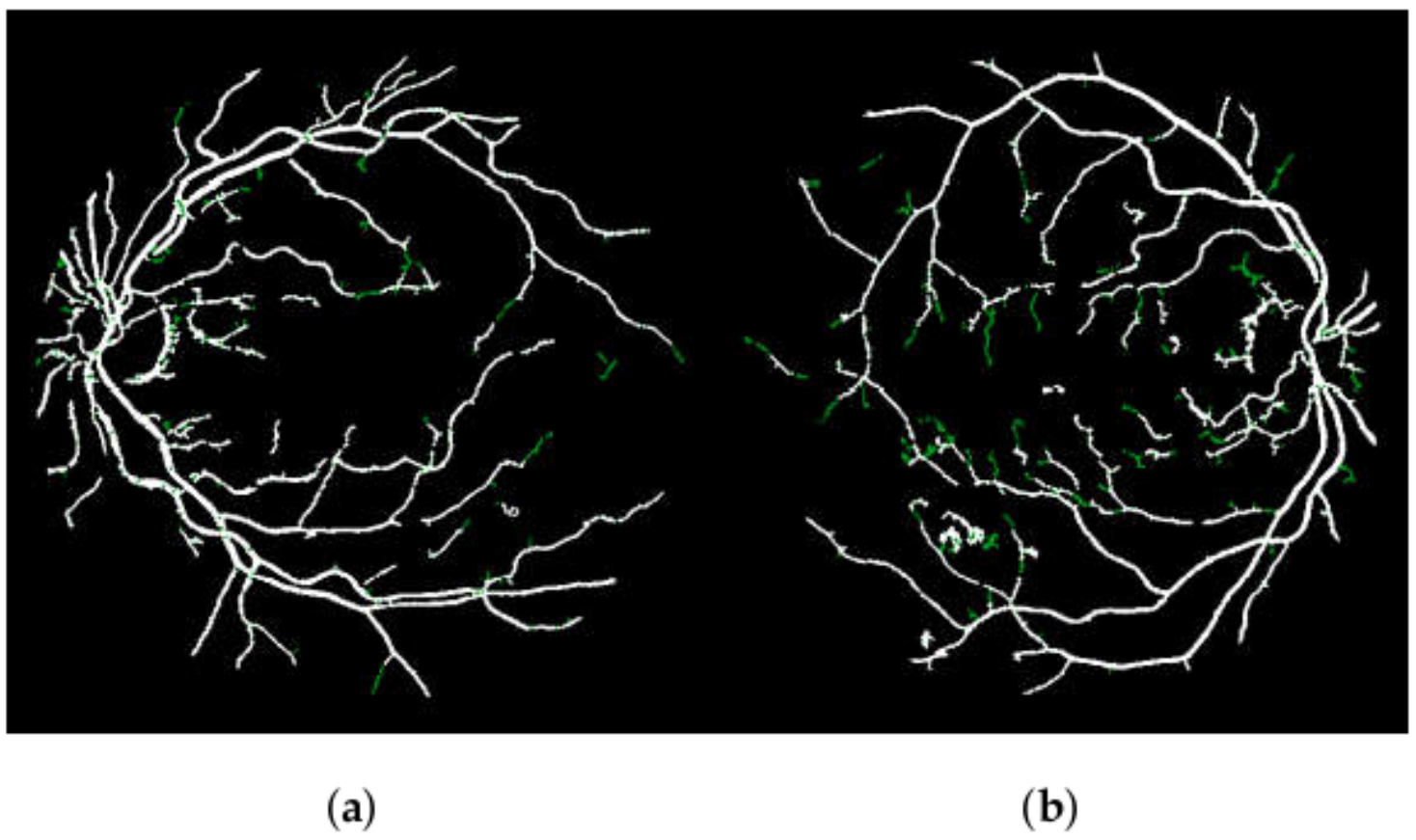

(**a**) (**b**)

**Figure 5.** Improvements obtained with the locally sensitive version of the connectivity filter (LS-CF). (**a**) Image 01 of the DRIVE dataset. (**b**) Image 14 of the DRIVE dataset.

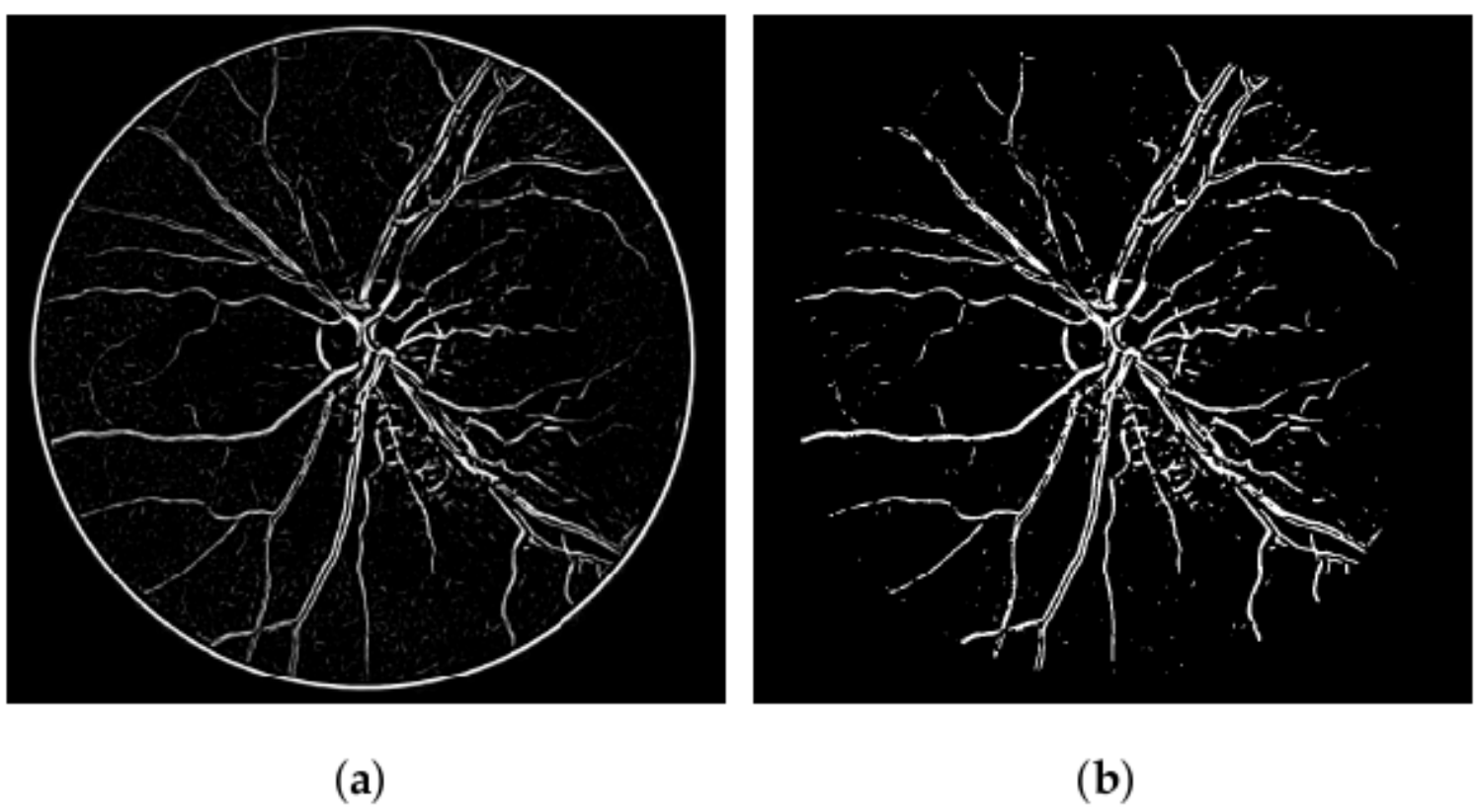

(**a**) (**b**)

**Figure 6.** *Cont.*

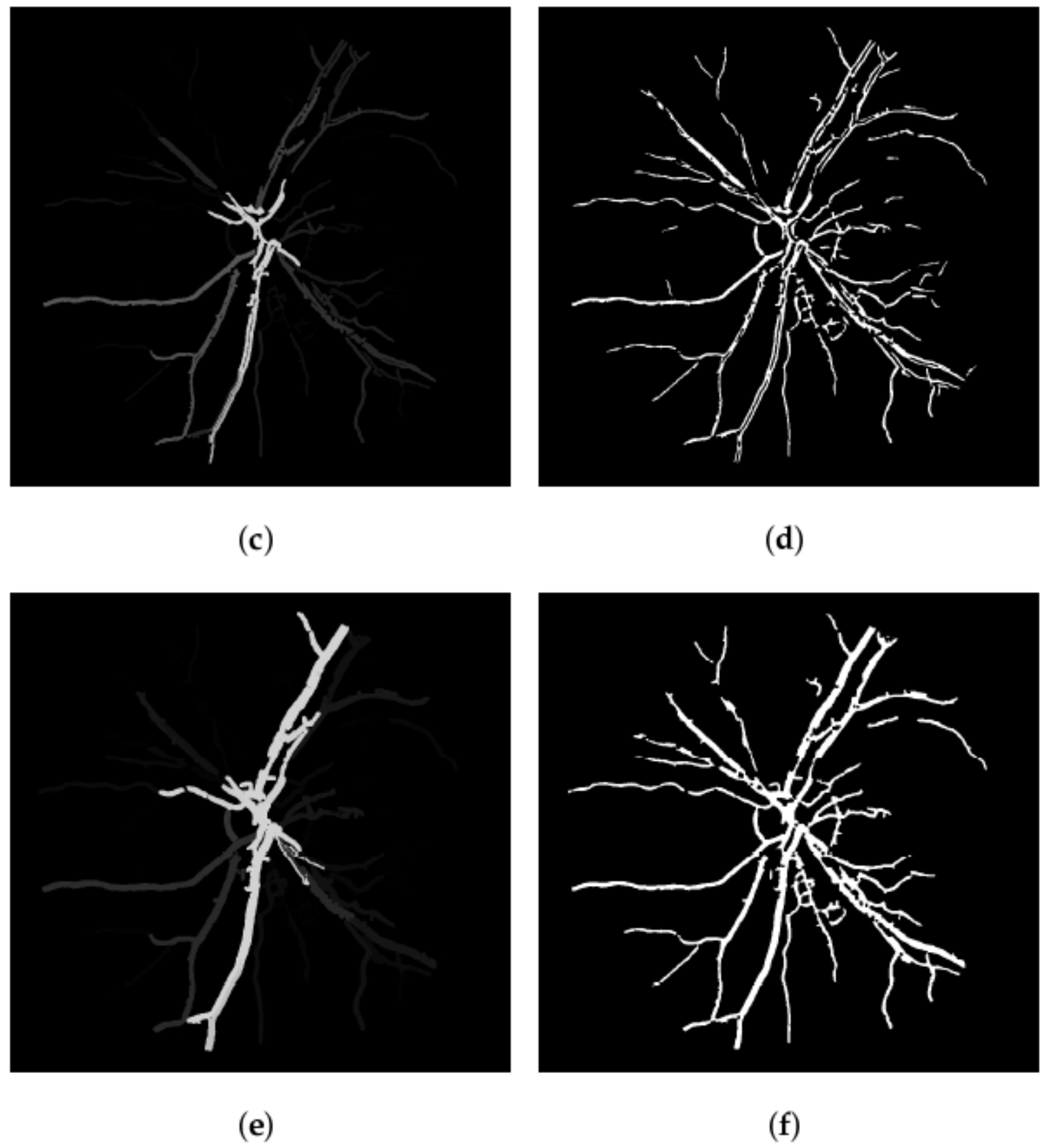

**Figure 6.** Steps and parameter analysis of the LS-CF algorithm. Image 14L of the CHASE-DB dataset. (**a**) The Frangi filter response. (**b**) The thresholded Frangi filter response shown in (**a**). (**c**) The LS-CF algorithm ($MAX_{score} = 350$ and $MAX_{dist} = 4$). (**d**) The thresholded version of the LS-CF response shown (**c**). (**e**) The LS-CF algorithm ($MAX_{score} = 350$ and $MAX_{dist} = 8$). (**f**) The thresholded version of the LS-CF response shown (**e**).

## 4. Performed Experiments

The experiments were performed on a total of five datasets that contain distinct image resolutions and a total of three different types of imaging modalities: retinal fundus images, laser ophthalmoscope images and cardiac X-ray angiography. The datasets used in this section are identified by their names, as follows: (1) the DRIVE dataset contains 20 retinal fundus images, (2) STARE also contains 20 retinal fundus images, (3) CHASE-DB contains 28 retinal fundus images, (4) IOSTAR contains 24 retinal scanner laser ophthalmoscope images and (5) OSIRIX contains 7 cardiac X-ray coronary images. More details about these datasets can be found in [4].

The method proposed in this work is unsupervised and therefore no training phase is required. The approach can segment vessels in different modalities without training. In terms of the experiments, for the datasets that have a folder separation (i.e., train and test), we calculated the results using the images in the test folder. For the datasets that do not originally have this separation proposed by their creators, we used the entire dataset.

Along with the CF and LS-CF proposals, we also evaluated a third possibility. This third possibility was created to measure and show that the local-sensitive connectivity filter is indeed more sophisticated than coupling the naive connectivity filter to a morphological closing, aiming to fill in the holes produced by the thresholded Frangi filter response and/or recreate vessel disconnections.

In this coupled approach, we use the morphological closing operation, which is composed of a morphological dilation followed by a morphological erosion [5]. The closing operation, by itself, is an operation used to remove holes from binary images and to reconnect structures. The used structuring element was the standard cross-shaped structuring element (distance of 1 pixel). Although the morphological approach can perform the reconnection of vessels to some extent, actual vessels can also be mistakenly connected to noise or non-vessel pixels.

The steps in this third approach are: (1) extract the green layer, (2) apply the Frangi filter, (3) threshold the result, (4) apply the connectivity filter, (5) threshold the result again and (6) apply the morphological closing. The closing operation does improve the results and disconnections in some cases, but it also increases the false positive rate in other situations, decreasing the overall accuracy. The mathematical morphology approach is not capable of achieving a solution at the same level of the result obtained with the LS-CF algorithm. We report this morphological approach in the tables of this section as "Connectivity Filter + Morphology Closing".

In terms of the visual results, Figure 7 compares the thresholded result from the Frangi filter with the result obtained with the proposed thresholded CF filter. It is clear that the CF removes the noise that originates from the Frangi response. The accuracy obtained with Figure 7 was 92.39% and the one obtained with the CF was 94.97% for this image.

The gray-level response of the connectivity filter is also interesting due to another property. This response shows the most important/core vessels (usually the thicker ones) of the retina. The response shown in Figure 7d is the response of the Frangi filter shown in Figure 7c after the application of the connectivity filter. The longer the vessel branch, the brighter the pixel color. This image is used to later generate the result in (e). The LS-CF response shown in (f) is an entirely different approach that changes how the connectivity score is computed, as shown in Algorithm 2.

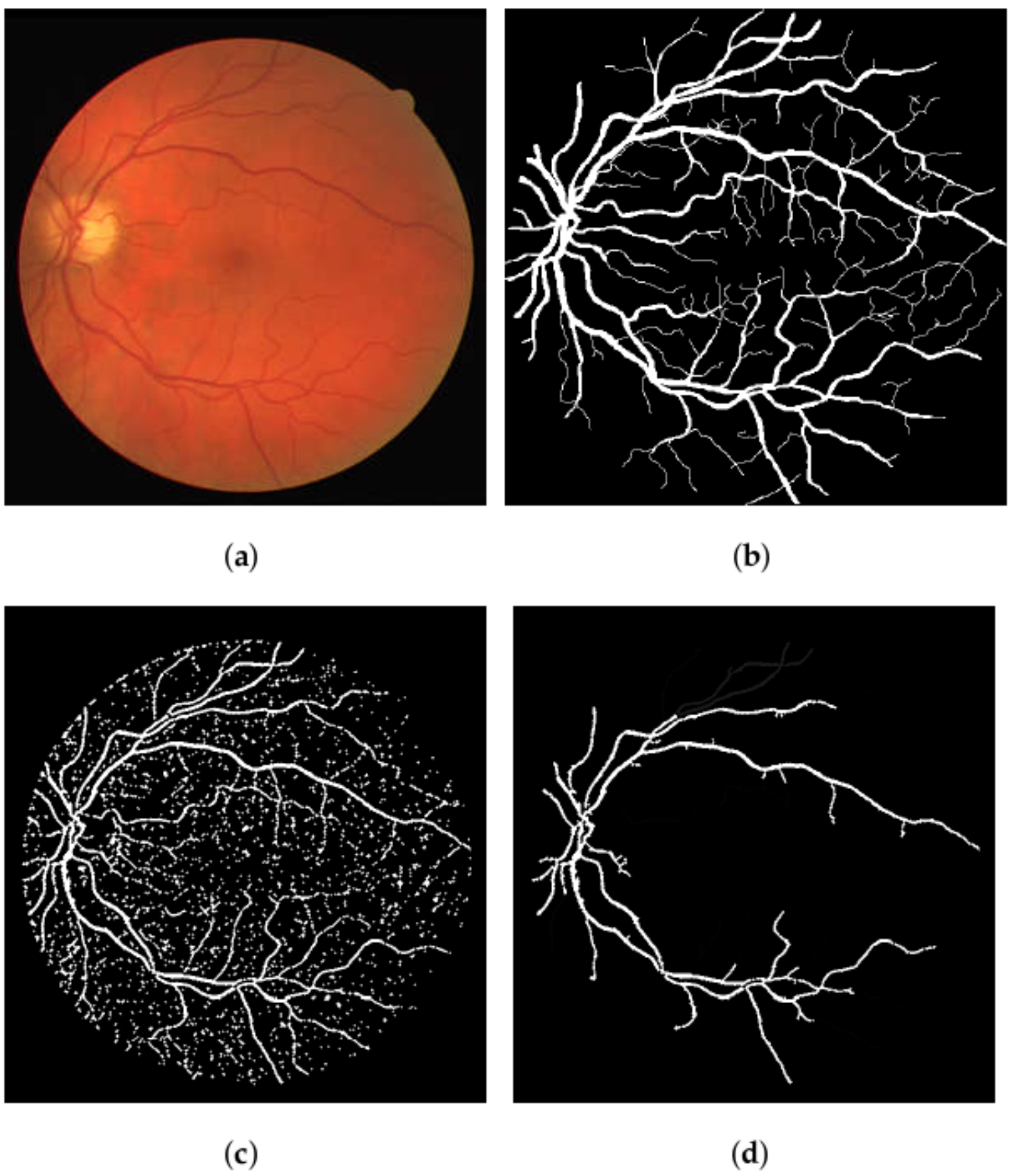

(a) (b)

(c) (d)

**Figure 7.** *Cont.*

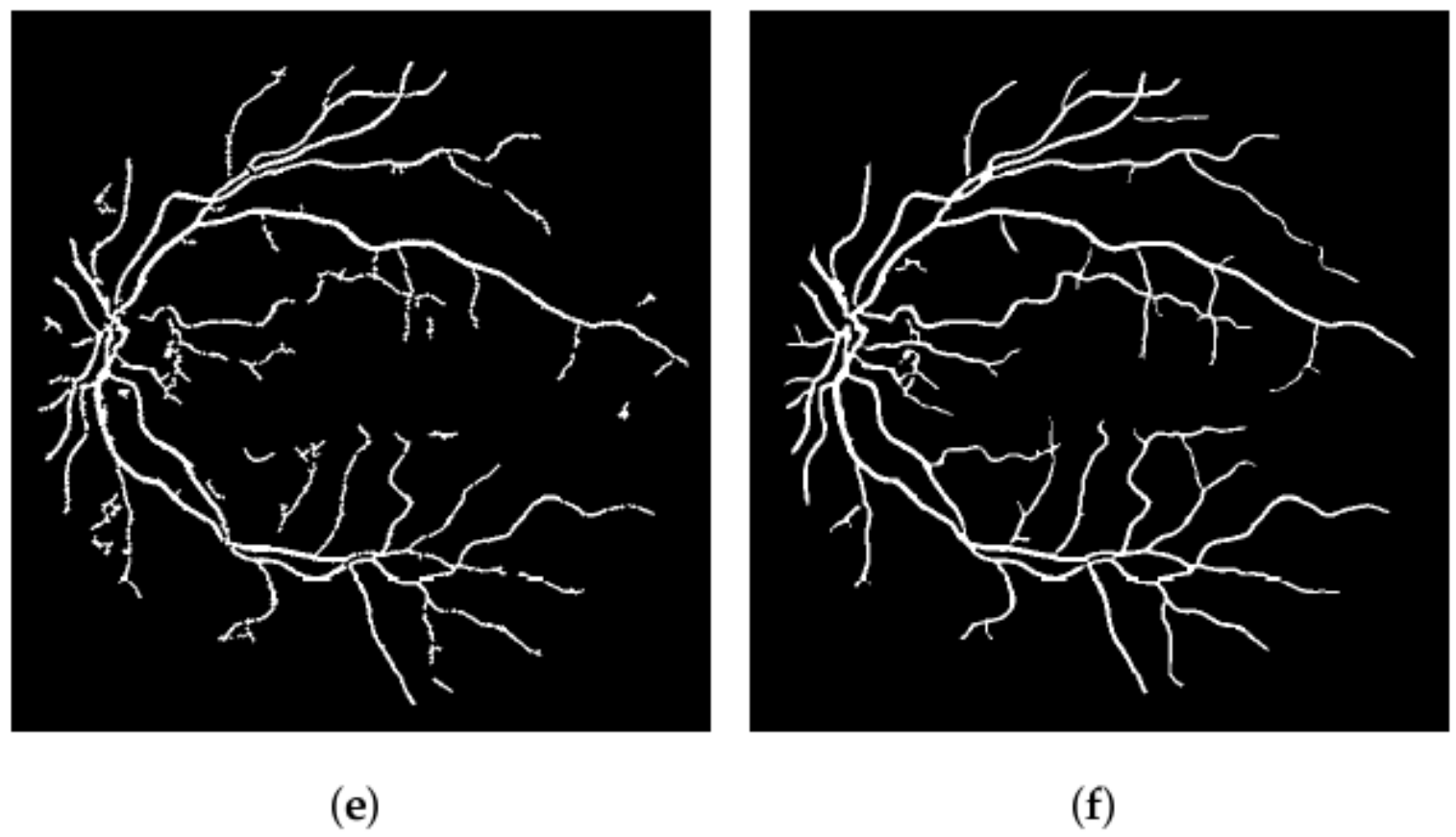

(e) (f)

**Figure 7.** Results using an image from the DRIVE dataset. (**a**) Image 05—DRIVE dataset. (**b**) Ground truth. (**c**) Thresholded Frangi. (**d**) Gray-level CF response (proposal). (**e**) Thresholded CF response (proposal). (**f**) LS-CF response (proposal).

The algorithm is able to achieve good results with the DRIVE dataset in general. No weird effect or strong segmentation error was observed. Image 08 is the only image that resulted in small circular artifacts, as shown in Figure 8 (out of the 20 images). This is one of the few unhealthy retinas in the DRIVE dataset. In this case, the Frangi filter response was able to achieve 92.22% accuracy, while the CF response achieved 92.63% and the LS-CF 92.60%, both improving the Frangi response.

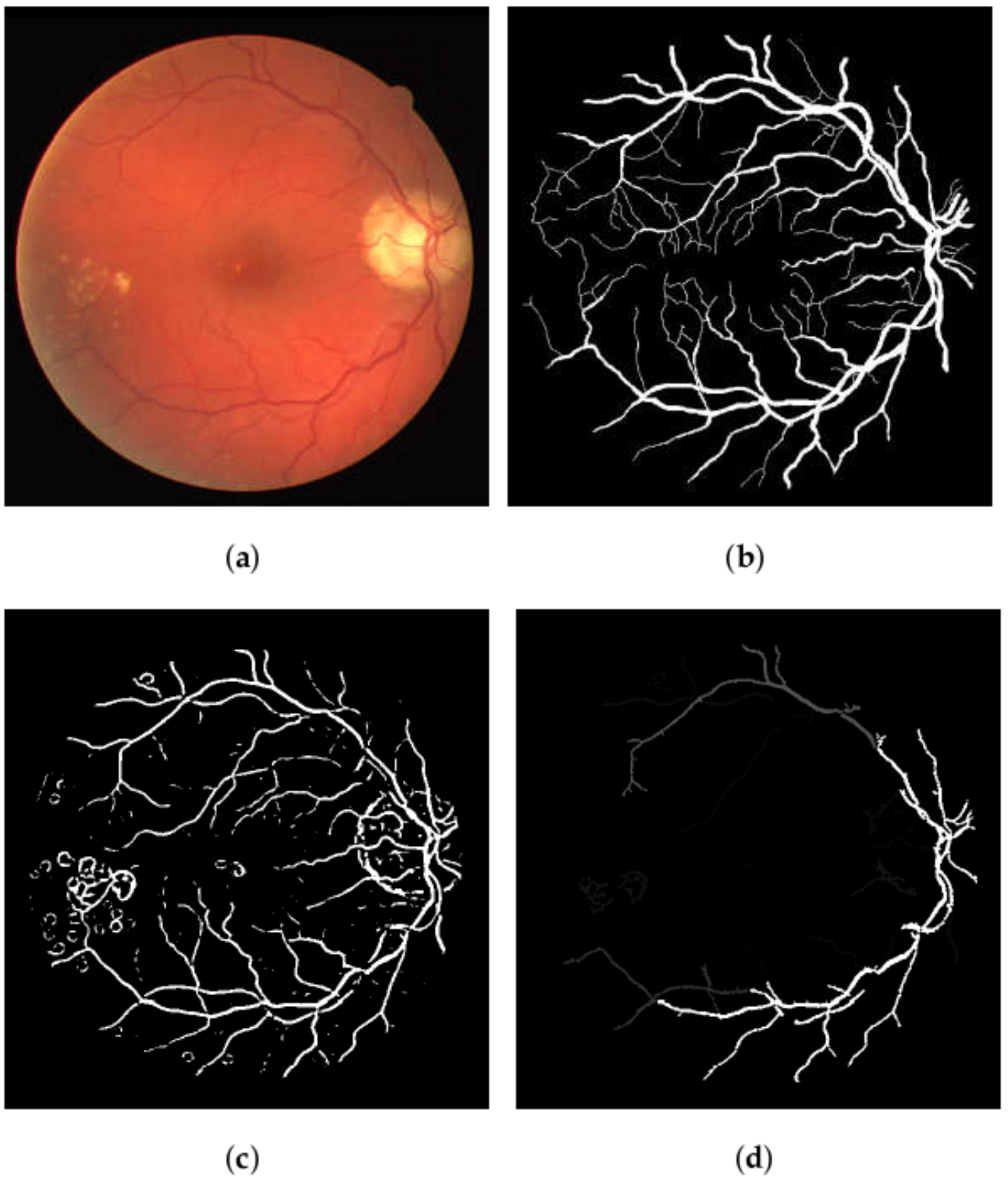

(a) (b)

(c) (d)

**Figure 8.** *Cont.*

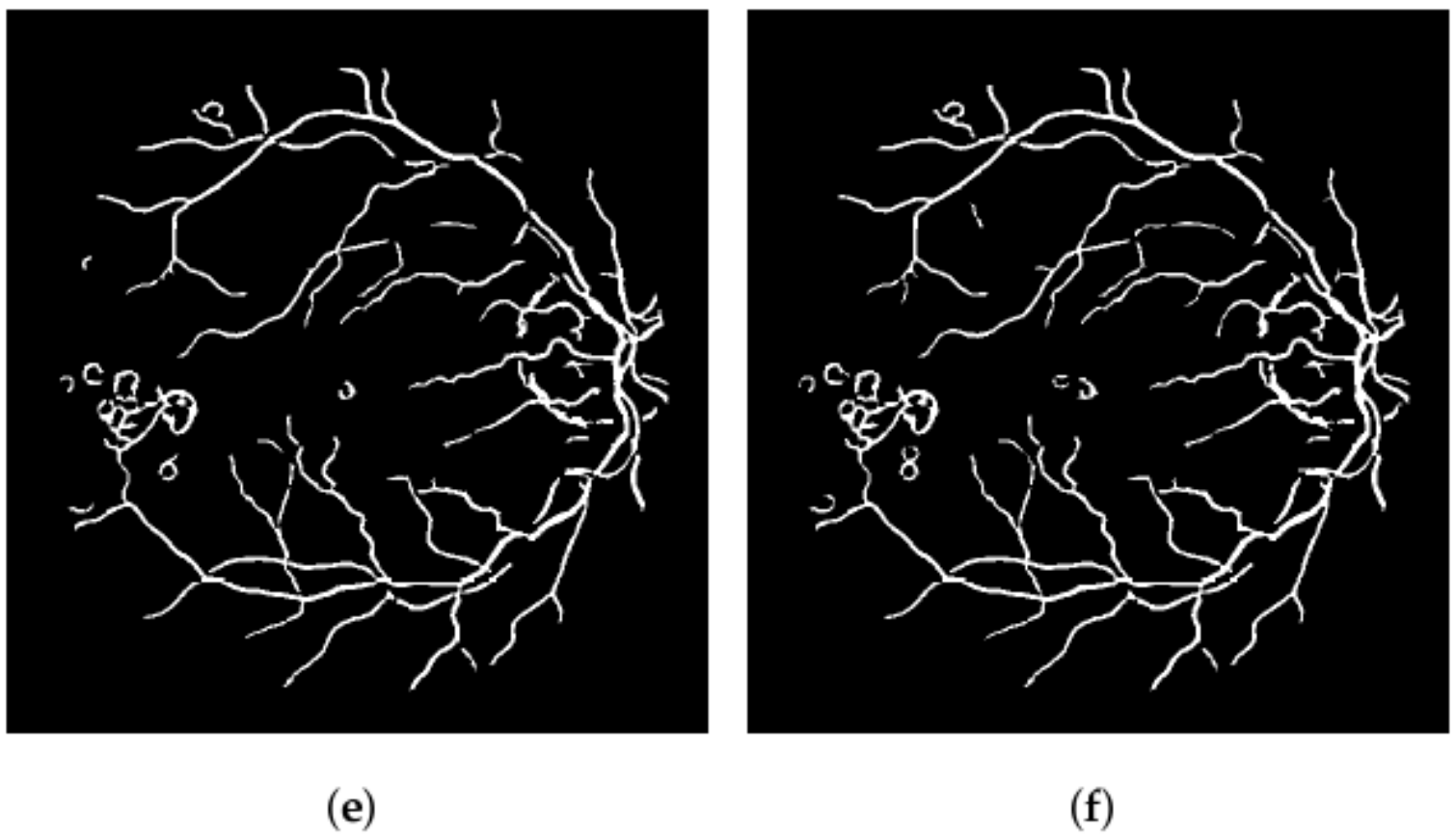

**Figure 8.** The worst numerical and visual segmentation result for the DRIVE dataset. (**a**) Image 08 from the DRIVE dataset. (**b**) Ground truth of image 08. (**c**) Thresholded Frangi response of image 08. (**d**) Gray-level CF response of image 08 (proposal). (**e**) Thresholded CF response of image 08 (proposal). (**f**) LS-CF response of image 08 (proposal).

Figure 9 shows the segmentation results of the LS-CF for a variety of datasets used in this work. The parameters should be adjusted when the modality is changed. However, these adjustments mostly relate to the response of the Frangi filter. In some cases, the Frangi should be adjusted to be more or less responsive to thinner or thicker vessels.

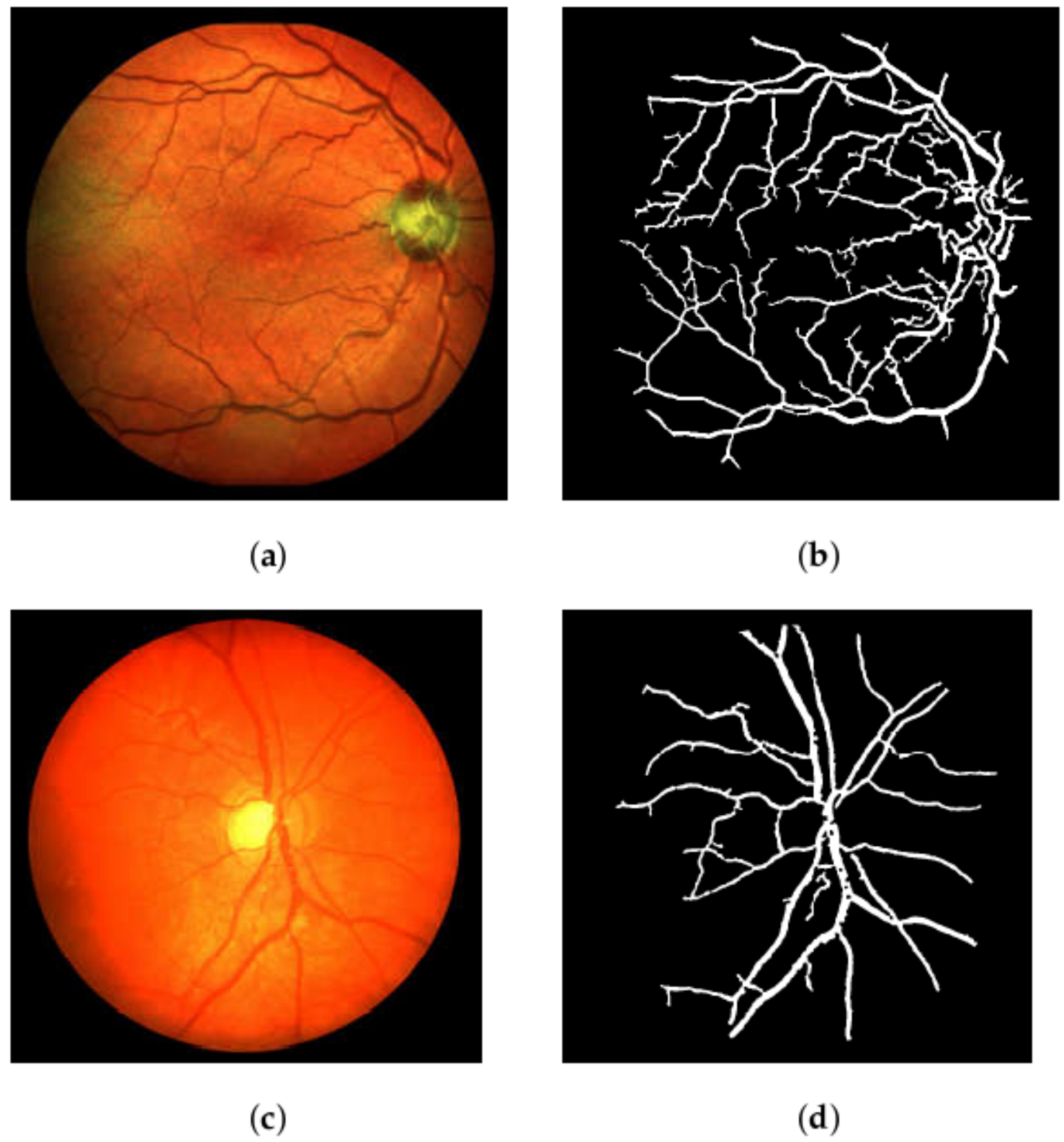

**Figure 9.** *Cont.*

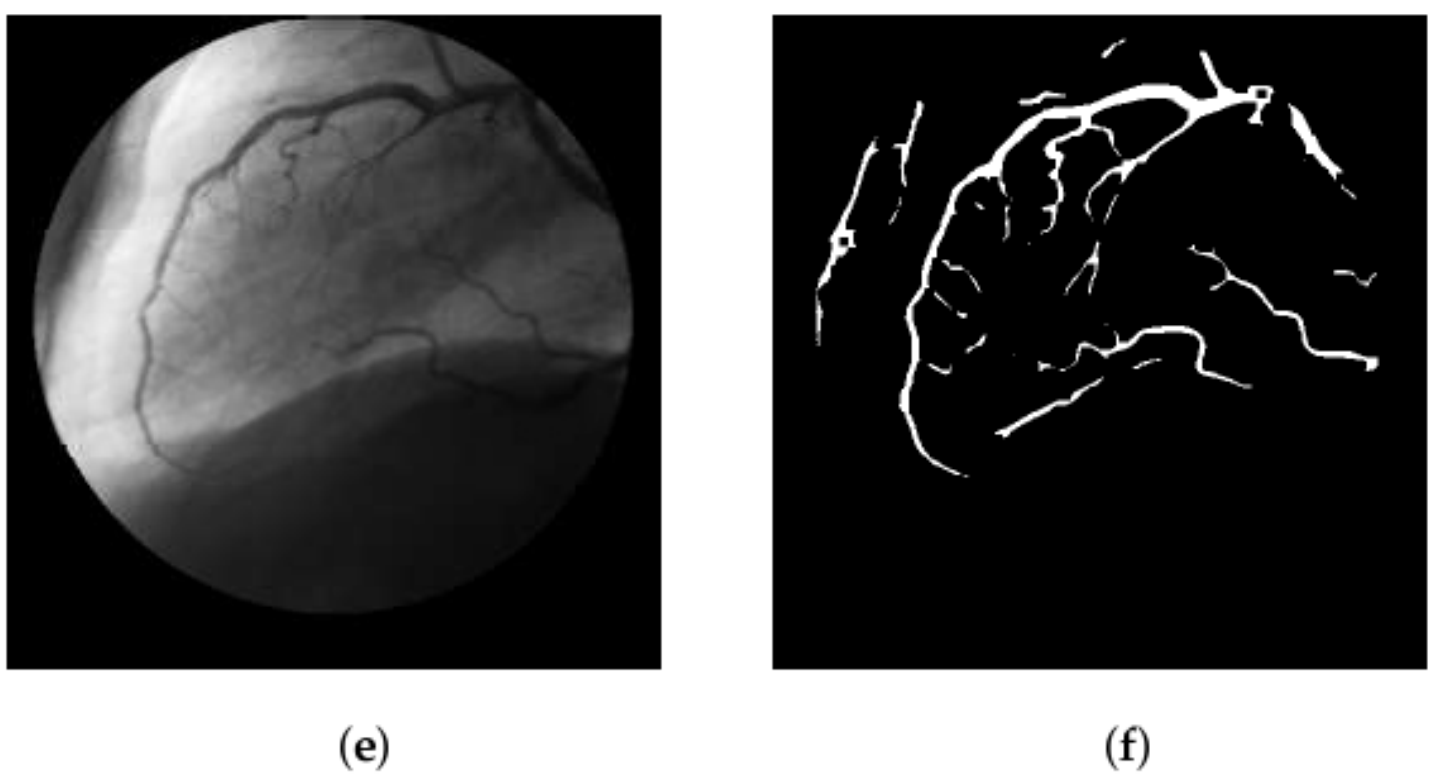

(e) (f)

**Figure 9.** Visual performance of the LS-CF in a variety of datasets. (**a**) Image 02 from the IOSTAR dataset (retinal scanner laser ophthalmoscope). (**b**) Segmentation result of image 02 from the IOSTAR dataset using the proposed LS-CF. (**c**) Image 11R from the CHASE-DB dataset. (**d**) Segmentation result of image 11R from the CHASE-DB dataset using the proposed LS-CF. (**e**) Image 4 from the OSIRIX X-ray angiographic dataset. (**f**) Segmentation result of image 4 from the OSIRIX X-ray angiographic dataset using the proposed LS-CF.

In what follows, we compare the numerical results obtained throughout the literature, a baseline method that consists of a thresholded Frangi filter response and three proposed approaches: (1) the connectivity filter, shown in Algorithm 1, (2) the naive connectivity filter + morphological closing to fill in the holes generated by the thresholded response and (3) the local-sensitive connectivity filter, shown in Algorithm 2.

Table 1 compares the results using the DRIVE dataset. In terms of the accuracy, the proposed LS-CF obtained an average of 95.77% accuracy and was able to outperform 12 out of 18 unsupervised methods. Table 2 shows the results obtained with the STARE dataset. In this case, the proposed approach was able to outperform 13 out of 14 unsupervised approaches.

**Table 1.** Comparison using the DRIVE dataset.

| Work | TP | TN | ACC |
|---|---|---|---|
| Supervised | | | |
| Staal et al., 2004 [31] | - | - | 94.41 |
| Sheet et al., 2013 [32] | - | - | 97.66 |
| Lupascu et al., 2010 [22] | - | - | 95.97 |
| Al-Rawi et al., 2007 [33] | - | - | 94.20 |
| Soares et al., 2006 [34] | - | - | 94.66 |
| Marin et al., 2011 [35] | 70.67 | 98.01 | 94.52 |
| Ricci et al., 2007 [36] | - | - | 95.95 |
| Maji et al., 2016 [37] | - | - | 93.27 |
| Li et al., 2016 [38] | 75.69 | 98.16 | 95.27 |
| Fu et al., 2016 [39] | 72.94 | - | 94.70 |
| Liskowski et al., 2016 [40] | 87.03 | 99.29 | 95.15 |
| Fan et al., 2016 [41] | 71.90 | 98.50 | 96.10 |
| Yan et al., 2018 [42] | 76.53 | 98.18 | 95.42 |
| Mo et al., 2017 [21] | 77.79 | 97.80 | 95.21 |
| Welikala et al 2017 [43] | - | - | 91.99 |
| Jin et al., 2019 [44] | 78.94 | 98.70 | 96.97 |
| Orlando et al., 2017 [45] | 78.97 | 96.84 | - |
| Jiang et al., 2019 [46] | 78.39 | 98.90 | 97.09 |
| Soomro et al., 2019 [47] | 87.00 | 98.5 | 95.60 |
| Adeyinka et al., 2019 [48] | 76.03 | - | 95.23 |
| Shin et al., 2016 [49] | 92.55 | 93.82 | 92.71 |
| Jebaseeli et al., 2019 [50] | 80.27 | 99.80 | 98.98 |
| Rodrigues et al., 2020 [4] | 89.83 | 98.16 | 97.4 |

**Table 1.** *Cont.*

| Work | TP | TN | ACC |
|---|---|---|---|
| Unsupervised | | | |
| Rawi et al., 2007 [33] | - | - | 94.20 |
| Lam et al., 2010 [51] | - | - | 94.72 |
| Zhang et al., 2010 [52] | 71.2 | - | 93.82 |
| Delibasis et al., 2010 [53] | 67.37 | 97.58 | 93.77 |
| Odstrcilik et al., 2013 [19] | 70.6 | 96.93 | 93.40 |
| Diri et al., 2009 [54] | 72.82 | 95.51 | - |
| Perez et al., 2010 [55] | 64.4 | - | 92.50 |
| Mendonca et al., 2006 [56] | 73.15 | - | 94.52 |
| Azzopardi et al., 2015 [57] | 76.55 | 97.04 | 94.42 |
| Zhang et al., 2016 [58] | 77.43 | 97.25 | 94.76 |
| Hossain et al., 2017 [59] | 78.63 | 97.11 | - |
| Srinidhi et al., 2018 [60] | 86.44 | 96.67 | 95.89 |
| Samant et al., 2019 [61] | 81.45 | 98.66 | 96.96 |
| Karn et al., 2018 [62] | 78.00 | 98.0 | 97.00 |
| Chakraborti et al., 2014 [63] | 72.05 | 95.79 | 93.70 |
| Memari et al., 2019 [25] | 76.1 | 98.01 | 96.1 |
| Tavakoli et al., 2021 [26] | 79.8 | 96.13 | 96.88 |
| Mahapatra et al., 2022 [27] | 70.2 | 98.44 | 96.05 |
| Baseline (Unsupervised) | | | |
| Thresholded Frangi Filter | 64.86 | 98.47 | 95.50 |
| Proposed Methods (Unsupervised) | | | |
| Connectivity Filter | 61.92 | 99.00 | 95.73 |
| Connectivity Filter + Morphology Closing | 64.81 | 98.69 | 95.70 |
| Local-Sensitive Connectivity Filter | 63.30 | 98.91 | 95.77 |

**Table 2.** Comparison using the STARE dataset.

| Work | TP | TN | ACC |
|---|---|---|---|
| Supervised | | | |
| Soares et al., 2006 [34] | - | - | 94.8 |
| Marin et al., 2011 [35] | - | - | 95.26 |
| Azzopardi et al., 2015 [57] | 77.16 | 97.01 | 94.97 |
| Ricci et al., 2007 [36] | - | - | 96.46 |
| Mo et al., 2017 [21] | 81.47 | 98.44 | 96.74 |
| Lupascu et al., 2010 [22] | - | - | 95.97 |
| Liskowski et al., 2016 [40] | 89.66 | 84.5 | 97.4 |
| Jin et al., 2019 [44] | 84.19 | 95.63 | 94.45 |
| Orlando et al., 2017 [45] | 76.8 | 97.38 | - |
| Ricci et al., 2007 [36] | - | - | 96.8 |
| Li et al., 2016 [38] | 70.27 | 98.28 | 95.45 |
| Fan et al., 2016 [41] | 70 | 97.9 | 95.9 |
| Jiang et al., 2019 [46] | 82.49 | 99.04 | 97.81 |
| Asad et al., 2015 [64] | 74.83 | 95.44 | 93.39 |
| Soomro et al., 2019 [47] | 84.8 | 98.6 | 96.8 |
| Adeyinka et al., 2019 [48] | 74.12 | - | 95.85 |
| Shin et al., 2016 [49] | 93.52 | 95.98 | 93.78 |
| Jebaseeli et al., 2019 [50] | 80.27 | 99.80 | 99.70 |
| Rodrigues et al., 2020 [4] | 94.26 | 98.62 | 98.27 |

**Table 2.** *Cont.*

| Work | TP | TN | ACC |
|---|---|---|---|
| Unsupervised | | | |
| Roychowdhury et al., 2015 [65] | - | - | 95.35 |
| Hoover et al., 2000 [66] | 80 | 90 | - |
| Mendonca et al., 2006 [56] | 67.64 | - | 94.79 |
| Lam et al., 2008 [67] | - | - | 94.74 |
| Annuziata et al., 2016 [68] | 71.28 | 98.36 | 95.62 |
| Zhao et al., 2015 [69] | 78 | 97.8 | 95.6 |
| Zhang et al., 2008 [70] | 93.73 | - | 90.87 |
| Jmani et al., 2015 [71] | 75.02 | 97.45 | 95.9 |
| Srinishi et al., 2018 [60] | 83.25 | 97.46 | 95.02 |
| Odstrcilik et al., 2013 [19] | 78.47 | 95.12 | 93.41 |
| Samant et al., 2019 [61] | 68.69 | 98.16 | 95.94 |
| Memari et al., 2019 [25] | 78.2 | 96.5 | 95.1 |
| Tavakoli et al., 2021 [26] | 79.8 | 95.89 | 96.46 |
| Mahapatra et al., 2022 [27] | 68.46 | 98.02 | 96.01 |
| Baseline (Unsupervised) | | | |
| Thresholded Frangi Filter | 68.84 | 97.37 | 95.26 |
| Proposed Methods (Unsupervised) | | | |
| Connectivity Filter | 63.23 | 98.73 | 96.06 |
| Connectivity Filter + Morphology Closing | 68.07 | 97.54 | 95.35 |
| Local-Sensitive Connectivity Filter | 65.41 | 98.62 | 96.09 |

The results obtained with the CHASE-DB dataset are shown in Table 3. The LS-CF outperformed all the unsupervised methods in the literature and outperformed two out of seven supervised approaches. This is the only case where the naive connectivity filter + morphology (morphological closing) obtained a slightly better accuracy when compared to the proposed LS-CF.

**Table 3.** Comparison using the CHASE-DB dataset.

| Work | TP | TN | ACC |
|---|---|---|---|
| Supervised | | | |
| Liskowski et al., 2016 [40] | 78.16 | 98.36 | 96.28 |
| Orlando et al., 2017 [45] | 72.77 | 97.12 | - |
| Jiang et al., 2019 [46] | 78.39 | 98.94 | 97.21 |
| Soomro et al., 2019 [47] | 88.6 | 98.2 | 97.60 |
| Adeyinka et al., 2019 [48] | 71.3 | - | 94.89 |
| Shin et al., 2016 [49] | 93.94 | 94.63 | 93.73 |
| Rodrigues et al., 2020 [4] | 87.82 | 98.52 | 97.78 |
| Unsupervised | | | |
| Srinishi et al., 2018 [60] | 82.97 | 96.63 | 94.74 |
| Samant et al., 2019 [61] | 70.27 | 98.28 | 95.45 |
| Chakraborti et al., 2014 [63] | 52.86 | 95.94 | 92.98 |
| Memari et al., 2019 [25] | 73.8 | 96.8 | 93.9 |
| Tavakoli et al., 2021 [26] | 79.8 | 95.12 | 94.75 |
| Baseline (Unsupervised) | | | |
| Thresholded Frangi Filter | 67.95 | 95.68 | 93.77 |
| Proposed Methods (Unsupervised) | | | |
| Connectivity Filter | 66.28 | 97.91 | 95.70 |
| Connectivity Filter + Morphology Closing | 71.21 | 97.83 | 95.96 |
| Local-Sensitive Connectivity Filter | 65.77 | 98.03 | 95.81 |

Table 4 compares the results using the IOSTAR dataset. Our approach was still able to outperform all works except our previous work that uses connectivity and machine learning—and is supervised.

Table 4. Comparison using the IOSTAR dataset.

| Work | TP | TN | ACC |
|---|---|---|---|
| Supervised | | | |
| Abbasi et al., 2015 [72] | 78.63 | 98.05 | 95.07 |
| Soares et al., 2006 [34] | 76.76 | 97.2 | 94.61 |
| Srinidhi et al., 2017 [73] | 88 | 84 | 89 |
| Zhao-li et al., 2018 [74] | 79.15 | 97.92 | - |
| Rodrigues et al., 2020 [4] | 86.49 | 98.96 | 98.04 |
| Baseline (Unsupervised) | | | |
| Thresholded Frangi Filter | 74.99 | 95.97 | 94.42 |
| Proposed Methods (Unsupervised) | | | |
| Connectivity Filter | 70.14 | 97.57 | 95.54 |
| Connectivity Filter + Morphology Closing | 71.3 | 97.06 | 95.16 |
| Local-Sensitive Connectivity Filter | 63.93 | 98.29 | 95.75 |

At last, Table 5 shows the results obtained with the OSIRIX dataset. In contrast to the previous comparisons, this is a comparison that is not performed using the same dataset, as previous articles in the literature did not publicly provide X-ray angiographic datasets. However, we are still able to measure how the LS-CF would most probably perform in these cases. We were also able to outperform all works in the literature.

Table 5. Comparison using the X-ray angiogram (OSIRIX) dataset.

| Work | TP | TN | ACC |
|---|---|---|---|
| Unsupervised | | | |
| Mhiri et al., 2013 [75] | 66 | - | - |
| Cervantes et al., 2016 [76] | - | - | 94.4 |
| Fatemi et al., 2010 [77] | - | - | 40.1 |
| Eiho et al., 1997 [78] apud Cervantes et al., 2016 [76] | - | - | 91.7 |
| Wang et al., 2012 [79] apud Cervantes et al., 2016 [76] | - | - | 93.1 |
| Kang et al., 2013 [80] apud Cervantes et al., 2016 [76] | - | - | 90.5 |
| Chanwimaluang et al., 2006 [81] apud Cervantes et al., 2016 [76] | - | - | 85.2 |
| Baseline (Unsupervised) | | | |
| Thresholded Frangi Filter | 50.51 | 95.56 | 94.41 |
| Proposed Methods (Unsupervised) | | | |
| Connectivity Filter | 46.16 | 96.42 | 94.91 |
| Morphology Closing + Connectivity Filter | 51.83 | 95.76 | 94.71 |
| Local-Sensitive Connectivity Filter | 46.26 | 96.43 | 94.93 |

## 5. Discussion

As shown in the results, the LS-CF is not equivalent to a CF coupled with a morphological closing. In fact, the LS-CF achieves the highest performance values in nearly all cases (with the only exception of Table 3). This is expected, as the usual mathematical morphology techniques are global techniques, i.e., they see and operate the entire image as a whole. In contrast to the LS-CF approach, this does not locally connect small vessel parts. The contrast here is similar to a local vs a global search [82].

Morphological closing consists of dilation plus erosion. In a practical view, the dilation fills some of the vessel disconnections. Later, however, when the erosion is subsequently applied, the disconnections are introduced again. This is expected as these disconnections are not quite literally "holes" in the image, and the closing operation works best in this case.

This problem of reintroducing the disconnections is only solved when more dilations are applied to the image (e.g., two dilations and one erosion). However, this thickens the vessel caliber and hence produces a lower accuracy and unrealistic visual segmentations. Often, it also introduces a lot of extra false positives to the results, as it begins to merge separated vessels together.

## 6. Conclusions

Automatic vessel segmentation is a challenging problem, and despite the amount of works in the literature, there is still room for improvement. Annotations provided by specialists are not always available, which denies the reproduction and utilization of supervised approaches. In some cases, they also need to be adjusted, i.e., by removing or including new features in order to provide an adequate response for a new modality. In contrast, unsupervised approaches are more adaptable for a vast amount of image modalities after a few adjustments of parameters and do not require training.

Although powerful and widely used, the Frangi filter still has its weaknesses, such as creating holes in the segmented vessels and discontinuities. In this work, we focus these issues raised by the Frangi filter and provide an enhancement to fill the discontinuities, improving accuracy and visual segmentations. Our approach is simple and just uses two filters (Frangi + LS-CF) to obtain the final result.

The proposed methodology consists of assigning vessel continuity scores to the pixels of a binary image (e.g., a thresholded Frangi). Longer connected components appear brighter. The connectivity score image is then thresholded to obtain the final result. The LS-CF algorithm introduces a tolerance to keep the algorithm traveling over the image pixels, attempting to re-link the vessel branches. Without the tolerance, the CF algorithm does not re-link the branches as expected. At the end, the small vessel components are deleted (usually noise) and we have nice, connected vessel branches.

This idea builds upon the hypothesis of the influence of each pixel to its surrounding pixels. Classical machine learning [83] is not strictly capable of using this connectivity data as a feature. In a previous work [4], we combined this connectivity idea to machine learning algorithms, producing some of the best results in the literature for vessel segmentation. However, in this new work, we create an unsupervised approach based on a very similar idea, showing that connectivity can be powerful enough to provide accurate segmentations.

The experiments were performed using a total of five datasets, where the proposed approaches obtained very competitive results in all cases. For the IOSTAR and OSIRIX datasets, our unsupervised approach was able to outperform all but our previous approach. When we analyze the unsupervised methods, our approach outperforms most of them in all dataset cases. In addition, the enhancements produced with our approach were visually better than the baseline thresholded Frangi response in all cases for all used datasets.

We did not perform an extensive tuning of the parameters in this work. For most cases, we visually tested some combinations and chose the one that seemed to be better for a few selections of images (usually 5 to 10). As a future work, a better optimization of parameters can be performed in order to enhance the obtained results as well as combining the proposed connectivity filters with machine learning, which would probably result in a very competitive methodology for vessel segmentation. A more robust pre-processing after the application of the Frangi and before the application of the CF and/or LS-CF can also improve the segmentation results further.

**Author Contributions:** Conceptualization, E.O.R., P.L.; methodology, E.O.R.; software, E.O.R., J.H.P.M.; validation, E.O.R., J.H.P.M., L.O.R.; formal analysis, E.O.R., J.H.P.M., M.T., D.C., J.T.O.; investigation, E.O.R., P.L., G.B.; resources, P.L., J.H.P.M., E.O.R.; data curation, E.O.R., J.H.P.M.; writing—original draft preparation, E.O.R., P.L.; writing—review and editing, E.O.R., D.C., M.T., J.T.O., L.O.R., G.B.; visualization, J.H.P.M.; supervision, E.O.R., D.C., M.T., J.T.O.; project administration, E.O.R., P.L.; funding acquisition, E.O.R., P.L., D.C., M.T., J.T.O. All authors have read and agreed to the published version of the manuscript.

**Funding:** This research received no external funding. The APC was partially funded by Universidade Tecnologica Federal do Parana-Pato Branco.

**Institutional Review Board Statement:** Not applicable.

**Informed Consent Statement:** Not applicable.

**Data Availability Statement:** Not applicable.

**Conflicts of Interest:** The authors declare no conflict of interest.